\documentstyle[fleqn]{article}

\renewcommand{\mathrm}{\rm}

\newcommand{\nl}{\nonumber \\}

\newcommand{\bq}{\begin{equation}}
\newcommand{\eq}{\end{equation}}
\newcommand{\ba}{\begin{eqnarray}}
\newcommand{\ea}{\end{eqnarray}}

\newcommand{\Rfcs}{${\cal R}$ functions}
\newcommand{\Rfc}{${\cal R}$ function}
\newcommand{\Li}{\mbox{Li}_{2}}
\newcommand{\be}{\begin{equation}}
\newcommand{\ee}{\end{equation}}
\newcommand{\itg}{\int \limits}
\newcommand{\eps}{\varepsilon}
\newcommand{\Ts}{\textstyle}
\newcommand{\barray}{\begin{array}}
\newcommand{\earray}{\end{array}}
\newcommand{\e}{\enspace}
\newcommand{\bdm}{\begin{displaymath}}
\newcommand{\edm}{\end{displaymath}}
\newcommand{\lslash}
           {\mbox{$ l \hspace{-0.9ex} \mbox{/} \hspace{-0.15ex} $}}
\newcommand{\qslash}
           {\mbox{$ q \hspace{-1.1ex} \mbox{/} \hspace{-0.05ex} $}}
\newcommand{\LLA}{\Large}

\begin{document}
\title{ A new Method for computing One-Loop Integrals }



\author{{L. Br\"ucher\footnote{(speaker)}, J. Franzkowski}\\
        {\small Institut f\"ur Physik, Johannes Gutenberg-Universit\"at,
         D-55099 Mainz, Germany}} 

\date{09/30/93}




\maketitle

\begin{abstract}

We present a new program package for calculating one-loop Feynman
integrals, based on a new method avoiding Feynman parametrization and
the contraction due to Passarino and Veltman. The package is
calculating one-, two- and three-point functions both algebraically and
numerically to all tensor cases. This program is written as a package
for Maple. An additional Mathematica version is planned later.

\end{abstract}




\section{Introduction \label{intr}}

In particle physics the calculation of one- and higher loop corrections
to particle processes is necessary to keep track with the increasing
accuracy of particle colliders. In this context it is very useful if a
fast and automatized calculation of the contributing Feynman diagramms
is provided by computer. On the most basic loop correction order, the
one-loop level, several attempts for such programs are already
available, like FF \cite{Ff} -- calculating scalar one-loop integrals
in Fortran -- or FeynCalc \cite{Mer} -- a Mathematica \cite{Math}
package -- which exhibits the usual Passarino-Veltman coefficients of
tensor integrals. Both programs are using the standard procedure for
calculating one-loop diagrams, namely Feynman parametrization, Wick
rotation and the Passarino-Veltman contraction for tensor structure
(cf. \cite{Hoo} and \cite{Pas}).

Although this standard procedure became very popular during the last
decade, it turned out to be very inconvenient for being implemented on
computer if the integral involves higher tensor structure. For that
reason we investigated in a new method for calculating arbitrary
one-loop integrals which is already published for the two- and
three-point case (cf. \cite{Kr2} and \cite{Kr3}).

Our approach is completely different from the standard procedure and
the structure of other packages. We are not forced to invert big
matrices which is the most difficult step in computing the tensor
structure of the integrals. In our method the loop momentum
integrations are performed directly, using the residue theorem and
introducing \Rfcs \ (see \cite{Car}), a class of special functions with
some nice and useful features. Instead of big matrices we get some
recursion relations for the involved \Rfcs \ (cf. \cite{Kr1} or
\cite{Fra}), which allow a faster and more memory efficient evaluation
of the integrals under consideration.

The program is designed to reduce the \Rfcs \ to a set of fundamental
functions which are rewritten in terms of logarithms and dilogarithms.
Up to now the different one-, two- and three-point functions are
programmed as procedures in Maple \cite{Map}. Therefore our results are
available numerically as well as algebraically. We tested our
procedures using Maple V on a VAX 4000-90 Workstation.

\section{Description of the method}

A detailed description of the method may be found in \cite{Kr1} or
\cite{Fra}. Here we would like to give a brief introductory overview
only. The calculation is based on the well-known dimensional 
regularization method, as described for example in \cite{Coll}. 
Instead of introducing Feynman parameters the integral is 
split up into orthogonal and parallel space (cf. \cite{Coll}),
so that for example the two-point integral takes on the form
\ba
\label{two}
B^{(p_{0} p_{1})}(q, m_{1}, m_{2}) = \nl \frac{2\pi^{\frac{D-1}{2}}}{\Gamma(\frac{D-1}{2})}
\itg_{-\infty}^{\infty} \!\! dl_{\|} \itg_{0}^{\infty} \!\! dl_{\bot} \, 
l_{\bot}^{D-2} \, 
\frac{(l_{\|})^{p_{0}} \, (l_{\bot})^{p_{1}}}
{{\cal P}_1 \, {\cal P}_2} 
\ea
with
\ba
 {\cal P}_1 = (l_{\|} + q)^{2} - l_{\bot}^{2} - m_{1}^{2} + i \varrho \nl
 {\cal P}_2 = l_{\|}^{2} - l_{\bot}^{2} - m_{2}^{2} + i \varrho \nonumber
\ea
Here we present the most general case, depending on one external
momentum $q$ and two arbitrary masses $m_{1}$ and $m_{2}$. The
splitting of space requires a distinction between $l_{\|}$ and
$l_{\bot}$ which we define as 
\ba \label{split1}
l_{\|} = \frac{l\cdot q}{\sqrt{q^2}} \qquad ; \qquad
l_{\bot} =  \sqrt{l_{\|}^2 - l^2}
\ea
In this notation $l_{\|}$ describes the component of $l$ which is parallel 
to the external momentum $q$, whereas $l_{\bot}$ represents the orthogonal 
complement.

The integrals (\ref{two}) can be reduced to a sum of \Rfcs \, by using 
(\ref{IntForm}) with the help of partial fraction and residue theorem.
The complete procedure is described in detail in \cite{Fra} and \cite{Kr1}. 
The result in the two-point-case may be written according to \cite{Kr2}:
\ba
B^{(p_{0} p_{1})}(q, m_{1}, m_{2})  =  i \pi^{\frac{3}{2} - \eps}
\, \frac{(-1)^{\Ts \frac{p_{1}}{2}} \, 
(e^{- i \pi})^{-\eps}}{q \, (- 2 \eps + p_{1} + 1)}  \nl
\qquad \times \frac{\Gamma(\frac{1}{2} + \eps - \frac{p_{1}}{2}) \, 
\Gamma(\frac{3}{2} -\eps + \frac{p_{1}}{2})}{\Gamma(\frac{3}{2} - \eps)} \nl
\times \left[ \, a_1 \, B(\eps - \frac{p_{1} + [p_{0}]}{2}, \frac{[p_{0}] + 
1}{2}) \, \times \right. \nl
 {\cal R}_{- \eps + \frac{p_{1} + [p_{0}]}{2}} \left({\Ts -
\frac{p_{1} + 1}{2} + \eps, 1; - m_{1}^{2} + i \varrho, -
(\frac{q}{2} + A)^{2}} \right) \nl
+ \sum_{i=0}^{p_{0}} \, {p_{0} \choose i} \, a_2
 \, B(\eps - \frac{p_{1} + [i]}{2}, \frac{[i] + 1}{2}) \, \times \\
\left. {\cal R}_{- \eps + \frac{p_{1} + [i]}{2}} 
\left({\Ts - \frac{p_{1} + 1}{2} + \eps, 1; - m_{2}^{2} + i \varrho, -
(\frac{q}{2} - A)^{2}} \right) \right] \nonumber
\ea
where $\eps = \frac{4 - D}{2}$ represents the usual dimensional 
regularization parameter. We used the abbreviations
\ba
a_1 = (-1)^{p_{0} - [p_{0}]} \,  (\frac{q}{2} + A)^{1 +
      p_{0} - [p_{0}]} \nl
a_2 = (-q)^{p_{0}-i} \,  (\frac{q}{2} - A)^{1 + i - [i]} \nl
A  = \frac{m_{2}^{2} - m_{1}^{2}}{2 q} \nl
\lbrack i] = \, \left\{ \e \barray{ll}
i, & \e \mbox{$i$ even} \\
i+1, & \e \mbox{$i$ odd}
\earray \right. \nonumber       
\ea
The result of the three-point integral is obtained by a quite similar 
procedure and may be found in \cite{Fra} or \cite{Kr3}.

\section{Evaluation and expansion of the \Rfcs}

After having expressed all possible integrals in terms of \Rfcs \ we 
get easy and fast computable results by applying some recurrence 
relations for the \Rfcs \ under consideration.

We achieve these relations by first reducing the number of different \Rfcs \ 
to a set of fundamental \Rfcs. This is possible due to the already 
mentioned features of the \Rfcs, especially the relation (\ref{ParInc}) 
and (\ref{AssFk}). The reduction takes place in two steps. First of all
we increase the parameter of the \Rfc \ to get it close to 0 or 1.
This can easily be achieved by using formula (\ref{ParInc}). In the second 
step we reduce the index of the \Rfc \ by virtue of a convenient formula
derived from (\ref{AssFk}). In the case of the two point-function it has 
the form:
\ba
{\cal R}_t(b_1,b_2;z_1,z_2) = \frac{-1}{t+\beta-1} \,\times \nl
\quad \lbrack \, ((t-1)z_1 z_2) \ {\cal R}_{t-2}(b_1,b_2;z_1,z_2) \nl
\qquad   + \, \left( (1-t-b_1)z_1 + (1-t-b_2)z_2 \right) \nl
\qquad \qquad \qquad \qquad \qquad \times \, {\cal R}_{t-1}(b_1,b_2;z_1,z_2) \rbrack 
\ea
Now the power of this method is obvious. Both steps mentioned above may be 
performed easily by computer, due to the recursive definition of the 
formulae (\ref{ParInc}) and (\ref{AssFk}).

Up to now we have acquired a result which is exact to all powers of $\eps$, 
though we are interested only in an expansion in powers of $\eps$. 
Therefore we now substitute the expansion for the fundamental \Rfcs \ 
in terms of $\eps$, resulting with simply computable functions like 
logarithms and dilogarithms. This is a rather cumbersome task but has to 
be performed only once for any N-point function. The procedure, especially 
its subtleties concerning analytical continuation, is described 
in detail in another paper \cite{BFK}. For completeness we 
here just cite the expansion for the scalar two-point case
\ba
 {\cal R}_{-\eps}(-{\Ts \frac{1}{2}}+\eps,1,z_1,z_2)  = 
   \frac{\pi i}{B(\eps,\frac{1}{2})} \,
       \frac{(z_2-z_1)^{\frac{1}{2}-\eps}}{z_2^{\frac{1}{2}}}  \nl
  \quad +  {\Ts \frac{1}{2}} \, z_1^{-\eps} \, \left(\frac{z_1}{  
  z_2}\right)^{1-\eps} \, \left[ \left(1+\sqrt{1-\frac{z_1}{z_2}}\right)^{-
1+2\eps} \right. \nl
\left. \qquad \qquad + \left(1-\sqrt{1-\frac{z_1}{z_2}}\right)^{-1+2\eps}
  \right] + {\cal O}(\eps^2)  
\ea

The three-point integrals
\ba \label{three}
C^{(p_{0} p_{1} p_{2})}(q_{1}, q_{2}, m_{1}, m_{2}, m_{3}) 
=  \frac{2\pi^{\frac{D-2}{2}}}{\Gamma(\frac{D-2}{2})} \times \\
\quad
\itg_{-\infty}^{\infty} \!\! dl_{0\|} \itg_{-\infty}^{\infty} \!\! dl_{1\|}
\itg_{0}^{\infty} \!\! dl_{\bot} \, l_{\bot}^{D-3} \,
\frac{(l_{0\|})^{p_{0}} \, (l_{1\|})^{p_{1}} \, (l_{\bot})^{p_{2}}}
{{\cal P}_1\,{\cal P}_2\,{\cal P}_3} \nonumber
\ea
\ba
{\cal P}_1 = (l_{0\|} + q_{1})^{2} - l_{1\|}^{2} - l_{\bot}^{2} - m_{1}^{2} + i 
               \varrho] \nl
{\cal P}_2 = (l_{0\|} + q_{20})^{2} - (l_{1\|} + q_{21})^{2} - l_{\bot}^{2} -
              m_{2}^{2} + i \varrho \nl
{\cal P}_3 = l_{0\|}^{2} - l_{1\|}^{2} - l_{\bot}^{2} - 
             m_{3}^{2} + i \varrho \nonumber
\ea
are calculated in a quite similar way (cf. \cite{Fra} or \cite{Kr3}).
Again we treat the general case with two independent external momenta
$q_{1},q_{2}$ and three masses $m_{1},m_{2},m_{3}$. For the momentum
components we write in a quite natural generalization:
\ba
l_{0\|} = \frac{l\cdot q_1}{\sqrt{q_1^2}} \nl
l_{1\|} = - \frac{l\cdot {q'}_2}{\sqrt{{q'}_2^2}}\,;\quad {q'}_2 = q_2 
               - \frac{q_1\cdot q_2}{q_1^2}\,q_1 \nl
\label{split2}
l_{\bot}  =  \sqrt{l_{0\|}^2 - l_{1\|}^2 - l^2} \\
q_{20}  =  \frac{q_1\cdot q_2}{\sqrt{q_1^2}} \nl
q_{21}  =  \sqrt{q_{20}^2 - q_{2}^2} \nonumber
\ea
The evaluation is resulting in the fundamental \Rfcs \
\ba 
{\cal R}_{-2\eps}(\eps, \eps, 1; x, y, z) =  \nl
 \  z^{-2\eps} \, \left\{ 1 \, + 
   \, 2\eps^2 \left[ \Li\left(1-\frac{x}{z}\right) + 
     \Lambda(x,z) \right. \right.\nl
\left. \left. \qquad \quad +  \Li\left(1-\frac{y}{z}\right) + \Lambda(y,z)     
       \right]  \right\} \, + \, O(\eps^3) \nl
{\cal R}_{-1-2\eps}(\eps, \eps, 1; x, y, z) =  (1-2\eps)z^{1-2\eps} \\
\  + \, \eps(x+y) \, + \,   
   2\eps^2 \left[ -x\ln x -y\ln y + z \, \Li\left(1-\frac{x}{z}\right) 
 \right. \nl
 \left. \quad +  z\, \Lambda(x,z) +  z\, \Li\left(1-\frac{y}{z}\right) 
    + z\, \Lambda(y,z) \right]   \, + \, O(\eps^3) \nonumber
\ea
\begin{table*}[hbt]
\setlength{\tabcolsep}{1.5pc}
\newlength{\digitwidth} \settowidth{\digitwidth}{\rm 0}
\catcode`?=\active \def?{\kern\digitwidth}
\caption{Input and Output for the one-loop procedures} \label{IO}
\begin{tabular}{ll}
\hline
input & output \\
\hline
OneLoop1Pt$(p,m)$ & $[\eps^{-1}\mbox{-{\em term}},\eps^0\mbox{-{\em term}}]$ \\
OneLoop2Pt$(p_0,p_1,q,m_1,m_2)$ & $[\eps^{-1}\mbox{-{\em term}},
\eps^0\mbox{-{\em term}}]$ \\
OneLoop3Pt$(p_0,p_1,p_2,q_{1},q_{20},q_{21},m_1,m_2,m_3)$
  & $[\eps^{-2}\mbox{-{\em term}},\eps^{-1}\mbox{-{\em term}},
     \eps^0\mbox{-{\em term}},\{abb.\}]$ \\
\hline
\end{tabular}
\\[0.4cm]
\end{table*}
here we introduced the $\Lambda$-Function,
\ba
 \Lambda(x,z)  \equiv 
 \ln \left(1 - \frac{x}{z}\right) \eta \left(x, \frac{1}{z}\right) \nl
  + \ln z \, \left[\eta\left(x - z,\frac{1}{1-z}\right)  
      - \eta\left(x - z,-\frac{1}{z}\right) \right]
\ea
Note that our relations are valid in all kinematical regions, so that 
the results of \cite{BFK} apply to all the cases of our program. For a 
detailed discussion see \cite{BFK}.

\section{The One-loop functions}

Our aim is to calculate one-loop integrals both algebraically and 
numerically. Therefore the algorithm is implemented in Maple \cite{Map}, 
a `language` for symbolic computations. To make the implementation for 
any user most comfortable, the procedures are bound to a package, which 
may be read from the Maple system during run-time. The user-interface 
consists of three elementary functions
\begin{itemize}
 \item OneLoop1Pt$(p,m)$

 \item OneLoop2Pt$(p_0,p_1,q,m_1,m_2)$

 \item OneLoop3Pt$(p_0,p_1,p_2,q_{1},q_{20},q_{21},m_1,m_2,m_3)$
\end{itemize}
The functions directly correspond to the integrals (\ref{two},\ref{three}).
For completness the one-point function
\ba
A^{(p)}(m) = \int \! d^{D}l \, \frac{(l)^{p}}{[l^{2} - m^{2} +
i \varrho]}
\ea
is implemented.

We followed again our notation which distinguishes between parallel and 
orthogonal space. The masses are denoted by $m_{i}$ and the definitions
of (\ref{split1}) and (\ref{split2}) for the momenta are of course
still valid. We would like to emphasize the fact that in our notation
the indices $p_i$ represent the powers of the different components of
the loop momentum $l$ and should not be mixed with Lorentz indices of
the corresponding tensor. This notation is perhaps not very often used,
but it corresponds directly to our method of integration. The usual
notation due to \cite{Pas} may be recovered by an additional algorithm 
also cointained in our package.

The results of the one-loop functions are Laurent expansions in terms of 
the ultraviolet regulator $\eps$. The output of the described procedures 
consists of a list where the significant coefficients of this expansion are 
denoted (see table \ref{IO}).

In the output ``$\eps^{-1}$-term'' is meant to represent the divergent
part (the coefficient of $\frac{1}{\eps}$) whereas ``$\eps^{0}$-term''
is describing the finite part. Of course the $\eps^{-2}$-term has to
vanish in any case, but is left in order to check the accuracy of the
program in numerical calculations.In the three-point case ``abb.'' is
written for some abbreviations which are introduced to get a more 
compact result.

As long as the arguments of the OneLoop functions are symbols the
output is given algebraically, whereas of course numbers inserted for
the arguments imply a numerical result. Moreover, in the numerical
case, the program sets the value of $\varrho$, the imaginary part of
the propagators, five digits higher than the numerical accuracy of the
whole calculation. In the algebraic case just 'rho' is returned.

For different calculations of the same OneLoop function it is of course
rather inconvenient to start the whole program again. Therefore our package
includes the functions OneLoopLib1Pt($n$), OneLoopLib2Pt($n$) and
OneLoopLib3Pt($n$) which generate a library of all OneLoop1Pt,
OneLoop2Pt and OneLoop3Pt functions respectively up to the tensor rank
$n$. Each function is written to one file. The filenames correspond to
the integrals in the following way:\\[1cm]
\begin{tabular}{ll}
ONE0.RES     & OneLoop1Pt$(0,m)$ \\
TWO00.RES    & OneLoop2Pt$(0,0,q,m_1,m_2)$ \\
TWO10.RES    & OneLoop2Pt$(1,0,q,m_1,m_2)$ \\
THREE000.RES & OneLoop3Pt$(0,0,0,q_{1},q_{20},$ \\
             & \qquad $q_{21},m_1,m_2,m_3)$ \\
THREE100.RES & OneLoop3Pt$(1,0,0,q_{1},q_{20},$ \\
             & \qquad $q_{21},m_1,m_2,m_3)$ \\
THREE010.RES & OneLoop3Pt$(0,1,0,q_{1},q_{20},$ \\
             & \qquad $q_{21},m_1,m_2,m_3)$ \\
$\cdots$     & $\cdots$
\end{tabular}
\\[1cm]
The path of the library may be specified with the variable
``tensorpath''.

\section{An example}

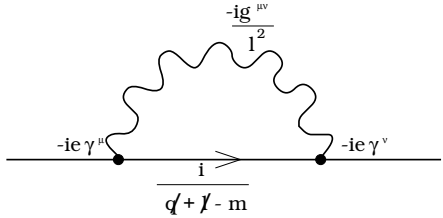
\begin{picture}(0,0)%
\includegraphics{eself.pstex}%
\end{picture}%
\setlength{\unitlength}{0.008750in}%
\begin{picture}(260,127)(60,670)
\end{picture}

To demonstrate how our package works we illustrate the calculation of the 
electron self-energy in QED. Starting with QED Feynman rules we get:
\ba
\Sigma (q) & = & -ie^2 \int \frac{d^Dl}{(2 \pi)^D} \, \gamma_{\mu} \,
   \frac{1}{\qslash + \lslash - m} \, \gamma_{\nu} \, \frac{g^{\mu \nu}}{l^2} 
   \nonumber\\
 & = & -ie^2 \int \frac{d^Dl}{(2 \pi)^D} \frac{\gamma_{\mu} (\qslash + 
  \lslash + m) \gamma^{\mu} }{[ (l+q)^2 - m^2 ] \, l^2} \nonumber \\
 & = & -ie^2 \int \frac{d^Dl}{(2 \pi)^D} \ \frac{1}{[(l+q)^2 - m^2 ] \, l^2} \nl
 &   & \times [ \gamma_{\mu} l_{\nu} \gamma^{\nu} \gamma^{\mu} +
          \gamma_{\mu} q_{\nu} \gamma^{\nu} \gamma^{\mu} + m \gamma_{\mu} 
          \gamma^{\mu} ] \nonumber \\
 & = & -ie^2 \int \frac{d^Dl}{(2 \pi)^D} \ \frac{1}{[(l+q)^2 - m^2 ] \, l^2}\nl
 &   &  \times [ l_{\nu} \gamma^{\nu}(2-D) + q_{\nu} \gamma^{\nu}(2-D) + mD  ] 
        \nonumber
\ea
Now we split into orthogonal and parallel space noting that the $\gamma$ 
matrix vector is split into $\gamma_{\|}$ and $\gamma_{\bot}$:
\ba
\Sigma (q) & = & -\frac{ie^2}{(2 \pi)^D} \{ (2-D)[ \gamma_{\|} \cdot B^{(10)}
              (q,m,0) \nl
   & &  - \gamma_{\bot} \cdot B^{(01)}(q,m,0) ] \nonumber \\
   & & + (2-D) q\cdot \gamma_{\|} B^{(00)}(q,m,0) \nl
   & & + D m B^{(00)}(q,m,0) \}  \nonumber
\ea
At this stage of calculation we now take the opportunity to incorporate our 
procedures which solve the ocurring integrals $B^{(p_0,p_1)}$. To be 
specific we have to compute the functions One\-Loop\-2Pt$(1,0,q,m,0)$,
OneLoop2Pt$(0,1,q,m,0)$ and OneLoop2Pt$(0,0,q,m,0)$. If we restrict 
ourselves in this small exercise to the divergent part ${\cal O}(\eps^{-1})$, 
we end with the following results:
\ba
\mbox{OneLoop2Pt}(1,0,q,m,0) \equiv B^{(10)}(q,m,0) = \nl
  - \frac{i \pi^2}{2 
             \eps} \, q \ + \ \frac{\pi^2}{4q^3}\, \Big\{ \cdots \Big\} \nl
\mbox{OneLoop2Pt}(0,1,q,m,0) \equiv  B^{(01)}(q,m,0)  = 0 \nl
\mbox{OneLoop2Pt}(0,0,q,m,0)  \equiv  B^{(00)}(q,m,0) = \nl
 \frac{i \pi^2}{\eps} 
 \ + \ \frac{\pi^{2}}{2 q^2}\, \Big\{ \cdots \Big\}   \nonumber
\ea
So our final result is
\ba
  \Sigma (q) & = & \frac{e^2}{16 \pi^2 \eps} (-q \cdot \gamma_{\|} + 4m) + 
               \mbox{finite} \nonumber \\
& = & \frac{e^2}{16 \pi^2 \eps} (-\qslash + 4m) + 
               \mbox{finite}
\ea
which is the well-known result for the divergent part of the electron
self-energy.

\section{Conclusion and further projects}

Our aim was to introduce a program package that gives any computer user
the possibilty to calculate one-loop particle processes without much
effort in manpower and hardware.

As a further project we plan to extend this package to the four- and
five-point function. Moreover we want to export our program package to
Mathematica \cite{Math}, because from our point of view Mathematica is
the more common language than Maple. Planned is also a menu-driven user
interface. But this might be rather difficult, because we haven't yet
found a convenient C-language package for algebraic calculations. (For
hints contact L. Br\"ucher)

\begin{appendix}

$\e$\\[0.5cm]
{\LLA \bf Appendix} \\

\setcounter{equation}{0}
\renewcommand{\theequation}{\mbox{A}.\arabic{equation}}

\subsection*{Conventions}

\begin{tabular}{ll}
   $D=4-2\eps$      &   Dimension in the dimensinal\\
                    & -regularisation-method \\
   $ {\cal R}_t(b_1,b_2,z_1,z_2)$  & \Rfc, t is called index, \\
                    &   $b_1,b_2$ parameters, \\
                    & $z_1,z_2$ arguments \\
   $ \beta=\sum_{i=0}^n b_i$ & sum of parameters of an \\
                    & \Rfc      \\  
 
\end{tabular}

\subsection*{Some fundamental properties of \Rfcs}

\begin{itemize}
\item formula for integrals
\ba \label{IntForm}
 \int \limits_0^{\infty} x^{\alpha-1} \, \prod_{i=1}^k (z_i+w_i x)^{-b_i} \, dx \nl =
  B(\beta - \alpha,\alpha) \, {\cal R}_{\alpha - \beta}({\bf b},{\bf \frac{z}{w} })
  \, \prod_{i=1}^{k} w_i^{-b_i}
\ea
\item increasing of parameters 
\ba \label{ParInc}
 \beta\,{\cal R}_t({\bf b},{\bf z}) & = &
        (\beta + t)\, {\cal R}_t({\bf b}+e_i,{\bf z}) \nl
  & &         \ -\ tz_i\, {\cal R}_{t-1}({\bf b}+e_i,{\bf z})
\ea
\item associated functions (k is the number of parameters)
\begin{equation}       \label{AssFk}
 \sum_{i=0}^k \ A_i^{(k)}(t,{\bf b},{\bf z})
            {\cal R}_{-t-i}({\bf b},{\bf z})\ =\ 0
\end{equation}
with
\ba                 \label{ADef}
 A_i^{(k)}(t,{\bf b},{\bf z}) = \frac{(t,i)(\beta-t-k,k-i)}{t(\beta-t-k)} \nl
 \qquad \qquad \times (t+i- \sum_{j=1}^k b_j z_j  \frac{\partial}{\partial z_j} ) E_i^{(k)}(z)
\ea                 
$E_i^{(k)} (z) $ is the elementary symmetric function, gained from
\be                \label{elemsym}
 (1+xz_1) \cdots (1+xz_k) = \sum_{i=0}^k x^i E_i^{(k)} (z)
\ee 

\end{itemize}

\end{appendix}


\end{document}